\documentclass[12pt]{article}
\usepackage{graphicx}
\usepackage[cp1251]{inputenc}
\begin{document}
\pagestyle{empty}

\pagestyle{headings} {\title{Influence of solar flares on behavior
of solar neutrino flux}}
\author{O.M.Boyarkin\thanks{E-mail:oboyarkin@tut.by}, G.G.Boyarkina\\
\small{\it{Belorussian State University,}}\\
\small{\it{Dolgobrodskaya Street 23, Minsk, 220070, Belarus}}}
\date{}

\maketitle
\begin{abstract}
Limiting ourselves to two flavor approximation the motion of the
neutrino flux in the solar matter and twisting magnetic field is
considered. For the neutrino system described by the 4-component
wave function $\Psi^T = (\nu_{eL}, \nu_{XL}, {\overline\nu}_{eR},
{\overline\nu}_{XL})$, where $X=\mu, \tau$, an evolution equation
is found. Our consideration carries general character, that is, it
holds for any SM extensions with massive neutrinos. The resonance
transitions of the electron neutrinos are investigated. Factors
which influence on the electron neutrino flux, crossing a region
of solar flares (SF) are defined. When the SF is absent a
terrestrial detector records the electron neutrino flux weakened
at the cost both of vacuum oscillations and of the MSW resonance
conversion only. On the other hand, the electron neutrino flux
passed the SF region in preflare period proves to be further
weakened in so far as it undergoes one (Majorana neutrino) or two
(Dirac neutrino) additional resonance conversions, apart from the
MSW resonance and vacuum oscillations.

The hypothesis of the $\nu_e$-induced decays which states that
decreasing the beta decay rates of some elements of the periodic
table is caused by reduction of the solar neutrino flux is
discussed as well.

PACS number(s): 12.60.Cn, 14.60.Pg, 96.60.Kx, 95.85.Qx, 96.60.Rd.
\end{abstract}

\section{Introduction}

The solar flares (SF) represents itself the most powerful of all
the solar activity events. The energy released during the SF is
about $10^{28}-10^{32}$ erg. It is now widely accepted that the
magnetic field provides a main energy source of the solar activity
including the SF's. Following observational results, theoretical
studies began to focus on the role of magnetic field in producing
the SF. A popular mechanism of the SF appearance is based on
breaking and reconnection of magnetic field strength lines of
neighboring spots (the magnetic reconnection model, for review,
see \cite{KST11}). At present this model has been considered to be
one of the promising mechanisms for producing the SF, although a
complete understanding of the relevant physics is still on the
way.

Observations suggest that the field strength $B_s$ of a big
sunspots ($d\sim2\times10^5$ km) could reach $10^4$ Gs while their
geometrical depth $h$ is approximately 300 km. Then the total
magnetic energy stored in such a sunspot with the volume $V=\pi
d^2h/4$ is
$$E_{mag}\simeq(B_s^2/8\pi)V\simeq4\times10^{34}\ \mbox{erg},\eqno(1)$$
which is sufficient to produce even the largest flare, although
only a small portion of this total energy can be used, that is, a
large amount of energy is unavailable because it is distributed as
the potential field energy. A magnetic field above sunspots is
characterized by geometrical phase $\Phi(z)$, and its first
derivative $\dot{\Phi}(z)$ where
$$B_x\pm iB_y = B_{\bot}e^{i\Phi(z)}\eqno(2)$$
(a coordinate system with the $z$-axis along the solar radius have
been chosen). A magnetic field above and under a spot has
non-potential character
$$(\mbox{rot}\ {\bf{B}})_z=4\pi j_z\neq 0\eqno(3)$$
(we are working in the natural system of units $\hbar=c=1$). The
data concerning centimeter radiation above a spot testify of a gas
heating up to the temperatures of a coronal order. Thus, for
example, at the height $\sim 2\cdot 10^2$ km the temperature
reaches the values of the order of $10^6$ K, which results in a
great value of solar plasma conductivity ($\sigma\sim T^{3/2}$).
That allows to suppose, that the density of longtitudal electric
current might be large enough in a region above a spot.

According to the magnetic reconnection model, a change of magnetic
field configuration in a sunspots group of fairly opposite
polarity might lead to the appearance of an limiting strength line
being common for whole group. Throughout the limiting line the
redistribution of magnetic fluxes takes place, which is necessary
for magnetic field to have the minimum energy. The limiting
strength line rises from photosphere to the corona. From the
moment of this line appearance an electric field induced by
magnetic field variations, causes current along the line, which
due to the interaction with a magnetic field takes a form of a
current layer. As the current layer prevents from the magnetic
fluxes redistribution, the process of magnetic energy storage of
the current layer begins. Duration of appearance and formation
period of the current layer (initial SF phase) varies from several
to dozens of hours. The second stage (an explosion phase of SF)
has a time interval of 1-3 minutes. At this stage magnetic energy
of sunspots transforms into kinetic energy of matter emission (at
a speed of $10^6\ m/s$ ), into energies of hard electromagnetic
radiation and into fluxes of solar cosmic rays (SCR) which consist
of protons $E_k\geq10^6$ eV of nuclei with charges $2\leq Z\leq28$
and energy within an interval from 0.1 to 100 eV/nucleon and of
electrons with $E_k\geq30$ MeV. SCR became a source, on the Sun
surface and later on the Earth atmospheres, of neutrons as well as
secondary kaons and pions. Their following secondary rays, as
muons $\mu^{\pm}$ and neutrinos and anti-neutrinos $\nu_{\mu},
\overline{\nu}_{\mu}$ as well as $\gamma$ rays, and their final
relic neutrinos $\nu_{\mu}, \overline{\nu}_{\mu}, \nu_e,
\overline{\nu}_e$ are also released by the chain reactions
$$\pi^{\pm}\to\mu^{\pm}+\nu_{\mu}(\overline{\nu}_{\mu}),
\qquad\pi^0\to2\gamma,\qquad \mu^{\pm}\to
e^{\pm}+\nu_e(\overline{\nu}_e)+\nu_{\mu}(\overline{\nu}_{\mu}).$$
The existence and detection of these neutrinos were first
predicted in Refs. \cite{JB88}.

The concluding stage (hot phase of SF) is characterized by
existence of high temperature coronal region and can continue for
several hours. The heating of dense atmospheric layers leads to an
evaporation of large amount of gas, which favors a long-continued
existence of a dense hot plasma cloud.

The high-power SF's can be especially destructive when they appear
to be aimed at the Earth, hitting the planet directly with
powerful charged particles. Such SF's are potentially dangerous
for satellites, power grids and astronauts. It is clear that the
prediction of the SF at the initial phase is a very important
task.

In 1995-1996 the series of works in which the correlation between
the SF's and solar neutrino flux have been published \cite{OBD95},
\cite{OB96},\cite{OB97}. So, for the first time one was supposed
to use the solar electron neutrinos for investigation of the solar
flares. Of course the detection of the neutrino flux correlation
with the SF will be possible only at the neutrino telescopes of
the next generation where events statistics will increase on
several orders of magnitude. However solving the problem comes
from the other hand, namely, from area of nuclear physics. In
recent years, a number of articles
\cite{JNJ09},\cite{EF09},\cite{DEK12},\cite{DO13} have been
published presenting evidence that some beta decay rates are
variable and this changeability may be connected with behavior of
the solar neutrino flux --- hypothesis of the $\nu_e$-induced
decays (see, for up-to-date review, Ref.\cite{TM16}). For example,
in 2006 J.Jenkins, monitoring a detector in his lab (1
$\mu\mbox{Ci}$ sample of $^{54}\mbox{Mn}$), discovered
\cite{JHJ2009} that the decay rate of\ $^{54}\mbox{Mn}$
$$^{54}\mbox{Mn}+e^-\to{^{54}\mbox{Cr}^*}+\nu_e\to{^{54}\mbox{Cr}}+
\gamma+\nu_e\eqno(4)$$ decreased slightly beginning 39 hours
before a large SF of 2006 Dec.13. Since then, researchers have
been examining similar variation in decay rates before SF's, as
well as those resulting from Earth's orbit around the Sun and
changes in solar rotation and activity. It should be noted that
the changeability of the decay rate has been observed only for
$\beta^{\pm}$ decay and electron capture processes.

The aim of this work is to demonstrate that, whether neutrino has
Dirac or Majorana nature, in the standard model (SM) extensions
decreasing the solar neutrino flux may occur in a preflare period.
This, in its turn, allows to explain the reduction of the decay
rate of some radioactive samples during the SF. In the second
chapter we consider a neutrino flux motion in solar matter. In so
doing we shall assume that the neutrino possesses the dipole
magnetic and anapole moments. The possible resonant transitions
which result in weakening the electron neutrino flux will be
found. The third chapter is devoted to discussion of the obtained
results.

\section{The resonant conversions of the solar neutrinos}

Let us find the evolution equation for the massive neutrinos in
two-flavor approximation ($\nu_e\nu_X$-mixing, $X=\mu,\tau$)
moving in the Sun. Not only do our consideration holds for SM
extensions having only the ordinary (light) neutrinos ($\nu_{eL},
\nu_{\mu L},\nu_{\tau L}$), but it also holds for SM extensions
having heavy neutrinos ($N_{eR},N_{\mu R},N_{\tau R}$) being
partners of the light neutrinos on the see-saw mechanism. The
heavy neutrinos appearing in some SM extensions are much more
heavier than the light neutrinos. For example, in the left-right
symmetric model the lower bound on the heavy neutrino mass is
approximately 100 GeV \cite{OM04}. As a result, these neutrinos do
not influence on oscillation picture of the solar neutrino whose
energy lies in the interval
$$0.14<E<14\ \mbox{MeV}.$$
As we are limited only by two generations, we should consider a
neutrino system consisting of $\nu_{eL}, \nu_{XL}$ and their
anti-particles $(\nu_{eL})^c, (\nu_{XL})^c$, where $c$ means an
operation of charge conjugation. It should be noted that Majorana
neutrino is also not an charge conjugation operator eigenstate due
to a switching on of weak interaction. As $(\nu_{eL})^c$ and
$(\nu_{eX})^c$ are right-handed neutrinos, in what follows we
shall use for them both in Majorana and Dirac cases following
notions ${\overline\nu}_{eL}$ and ${\overline\nu}_{XL}$
respectively.

For the case of the Majorana neutrino nature the evolution
equation in a Schrodinger-like form takes the form
$$i\frac{d}{dz}\left(\matrix{\nu_{eL}\cr
\nu_{XL}\cr\overline{\nu}_{eL}\cr\overline{\nu}_{XL}\cr}\right)
={\cal H}\left(\matrix{\nu_{eL}\cr
\nu_{XL}\cr\overline{\nu}_{eL}\cr\overline{\nu}_{XL}\cr}\right)
,\eqno(5)$$ where
$${\cal H}=\left(\matrix{{\cal H}_{\nu\nu}& {\cal H}_{\nu\overline{\nu}}\cr
           {\cal H}^{\dagger}_{\nu\overline{\nu}}   & {\cal
           H}_{\overline{\nu}\
           \overline{\nu}}\cr}\right),$$
           $${\cal H}_{\nu\nu}=\left(\matrix{
           \delta^{12}_c+V_{eL}+4\pi a_{\nu_e\nu_e}j_z &
           -\delta^{12}_s+4\pi a_{\nu_e\nu_X}j_z\cr
           -\delta^{12}_s+4\pi a_{\nu_X\nu_e}j_z
           &-\delta^{12}_c+V_{XL}+
           4\pi a_{\nu_X\nu_X}j_z \cr}\right),$$
$$\delta^{12}_{c(s)}=\frac{m^2_1-m^2_2}{4E}\cos{2\theta_{\nu}}
(\sin{2\theta_{\nu}}),\qquad V_{eL}=\sqrt{2}G_F(N_e-N_n/2),$$
$$V_{XL}=-\sqrt{2}G_FN_n/2,\qquad
{\cal H}_{\nu\overline{\nu}}=\left(\matrix{ 0 &
\mu_{\nu_e\overline{\nu}_X} B_{\bot}e^{i\Phi} \cr
-\mu_{\nu_e\overline{\nu}_X}B_{\bot}e^{i\Phi} & 0 \cr}\right),$$
$${\cal H}_{\overline{\nu}\ \overline{\nu}}={\cal
H}_{\nu\nu}(V_{lL}\to-V_{lL},j_z\to-j_z),$$ $i$- and $k$-neutrino
states, $N_e$ and $N_n$ are electron and neutron densities,
respectively, $\theta_{\nu}$ is a mixing angle in vacuum between
mass eigenstates $\nu_1$ and $\nu_2$, $V_{eL}$ ($V_{XL}$) is a
matter potential describing interaction of the $\nu_{eL}$
($\nu_{XL}$) neutrino with a solar matter. The SM extensions with
the extra gauge groups $SU(2)$ or $U(1)$ predict additional gauge
bosons whose low bounds on masses lie in the region of 3 TeV and
above. Therefore, when calculating the matter potential one may
neglect their contribution to the matter potential.

When the neutrino is a Dirac particle
${\cal{H}}_{\nu\overline{\nu}}$ should be replaced by the
expression
$${\cal{H}}_{\nu\overline{\nu}}=\left(\matrix{\mu_{\nu_e\overline{\nu}_e}
B_{\bot}e^{i\Phi} & \mu_{\nu_e
\overline{\nu}_X}B_{\bot}e^{i\Phi}\cr
\mu_{\nu_e\overline{\nu}_X}B_{\bot}e^{i\Phi} &
\mu_{\nu_X\overline{\nu}_X}B_{\bot}e^{i\Phi}\cr}\right)\eqno(6)$$
and assume $V_{lL}$ equal to zero in the expression for ${\cal
H}_{\overline{\nu}\ \overline{\nu}}$.

Before proceeding further we discuss the experimental bounds on
the neutrino multipole moments (MMs) and compare them with
theoretical predictions. The most sensitive and established method
for the experimental investigation of the neutrino MMs is provided
by direct laboratory measurements of (anti)neutrino-electron
elastic scattering in solar, accelerator and reactor experiments.
A detailed description of such experiments could be found in Ref.
\cite{bed07}.

Let us take up first the dipole magnetic moments (DMMs) for Dirac
neutrinos. The analysis of the recoil electron spectrum in the
SuperKamiokande experiment gave \cite{DVL04}
$$\mu_{\nu}\leq1.1\times10^{-10}\mu_B.\eqno(7)$$
An upper limit on the neutrino DMM
$\mu_{\nu}\leq8.5\times10^{-11}\mu_B$ which was found in an
independent analysis of the first stage of the Borexino experiment
\cite{Dmon08} results in the following bounds for $\nu_{\mu}$ and
$\nu_{\tau}$
$$\mu_{\nu_{\mu}}\leq1.5\times10^{-10}\mu_B,\qquad
\mu_{\nu_{\tau}}\leq1.9\times10^{-10}\mu_B.\eqno(8)$$ The neutrino
interaction with the solar magnetic field could lead to the
resonant conversion $\nu_e\to\overline{\nu}_e$. Using this effect
produces the inequality \cite{Gon}
$$\mu_{eff}(\nu_{{}^8B})\leq(10^{-10}\div10^{-12})\mu_B.\eqno(9)$$
At the moment the world best limit on electron neutrino DMM is
coming from the GEMMA experiment at the Kalinin nuclear power
plant \cite{AGB12}
$$\mu_{\nu_e}\leq2.9\times10^{-11}\mu_B\qquad (90\%\mbox{C.L.}).\eqno(10)$$
Note that the bounds on transit DMMs shall be obtained under
observation of the processes
$$\nu_l+e^-\to \nu_{l^{\prime}}+e^-,\qquad
\overline{\nu}_l+e^-\to \overline{\nu}_{l^{\prime}}+e^-, \qquad
(l\neq{l^{\prime}})\eqno(11)$$ which proceed with the partial
lepton flavor violation.

As far as a Majorana neutrino is concerned, the global fit of the
reactor and solar neutrino data gives the following values for
transition DMMs \cite{WGri03}
$$\mu_{12},\mu_{13},\mu_{23}\leq1.8\times10^{-10}\mu_B.\eqno(12)$$

The theoretical predictions of the minimally extended SM (MESM)
are very far from upper experimental bounds \cite{BWLe77}
$$\mu_{\nu_i}=3.2\times10^{-19}\mu_B\left({m_{\nu_i}\over1\
\mbox{eV}}\right),\eqno(13)$$ and
$$\mu_{\nu_i\nu_i^{\prime}}\approx10^{-4}\mu_{\nu_i}.\eqno(14)$$ So, in the
MESM case the neutrino DMMs are negligibly small and are of no
physical interest. On the other hand in alternative SM extensions
the neutrino DMMs may have the values close to the experimental
bounds. One such SM modification is the model in which, along with
the light right-handed neutrinos, a charged scalar $\eta^{(\pm)}$
being $SU(2)_L$ singlet is introduced into a theory \cite{AZe85}.
In this model the neutrino DMM could be as large as
$10^{-11}\mu_B$. Similar value is also predicted by the left-right
symmetric model \cite{OMG14}.

We are coming now to the discussion on the anapole moment (AM). At
neutrino mass neglecting the AM is associated with a neutrino
charge radius (NCR) by the relation
$$a_{\nu_l}={1\over6}<r_{\nu_l}^2>.\eqno(15)$$
Within the MESM the gauge-invariant result for the NCR has been
obtained \cite{JBLG}
$$<r_{\nu_l}^2>={G_F\over4\sqrt{2}\pi^2}\Bigg[3-2\log\Bigg(
{m_l^2\over m_W^2} \Bigg)\Bigg].\eqno(16)$$ It gave, in turn, the
following numerical value
$$<r_{\nu_l}^2>=4\times10^{-32}\ \mbox{cm}^2.\eqno(17)$$
Note that the NCR can be treated as an effective scale of the
particle's size, which should influence physical processes such
as, for instance, elastic neutrino-electron scattering. Then in
order to take into account the contribution coming from the NCR to
the cross section of this process the following substitution can
be done
$$g_V\to{1\over2}+2\sin^2\theta_W+{2\over3}m_W^2<r_{\nu_l}^2>
\sin^2\theta_W.\eqno(18)$$ Using this scheme, the TEXONO
collaboration found \cite{TEXONO}
$$-2.1\times10^{-32}\ \mbox{cm}^2<<r_{\nu_l}^2>3.3\times10^{-32}\
\mbox{cm}^2\qquad (90\%\mbox{C.L.})\eqno(19)$$ There are other
limits on the electron neutrino charge radius as well. They are
obtained:\\
from neutrino neutral-current reactions \cite{RA91}
$$-2.74\times10^{-32}\ \mbox{cm}^2<r_{\nu_e}^2>4.88\times10^{-32}\
\mbox{cm}^2\qquad (90\%\mbox{C.L.}),\eqno(20)$$ from solar
experiments (Kamiokande II and Homestake) \cite{AMM92}
$$<r_{\nu_e}^2><2.3\times10^{-32}\ \mbox{cm}^2\qquad (95\%\mbox{C.L.},
\eqno(21)$$ from an evaluation of the weak mixing angle
$\sin^2\theta_W$ by a combined fit of all electron neutrino
elastic scattering data \cite{JB08}
$$-0.13\times10^{-32}\ \mbox{cm}^2<<r_{\nu_e}^2>3.32\times10^{-32}\
\mbox{cm}^2\qquad (90\%\mbox{C.L.}).\eqno(22)$$ The effects of new
physics beyond the SM can also contribute to the NCR (see, for
example, \cite{HNS08}).

Now we return to the evolution equation (5). Here one should get
rid of an imaginary part in a Hamiltonian. It could be done by the
transition to reference frame (RF), rotating at the same angle
speed as a magnetic field. The expression for Hamiltonian in this
RF follows from the initial one by the following substitution
$$e^{\pm i\Phi}\to1,\qquad V_{lL}\longrightarrow V_{lL}-
\frac{\dot{\Phi}}{2}.\eqno(23)$$ Let us discuss the possible
resonance conversions only for left-handed electron neutrino. In
the case of the Majorana neutrino we have: \\
(i) $\nu_{eL}\to\nu_{XL}$ is the so-called
Micheev-Smirnov-Wolfenstein (MSW) resonance, which is realized if
the condition $$\Sigma_{\nu_{eL}\to\nu_{XL}}=
2\delta^{12}_c+V_{eL}-V_{XL}+4\pi (a_{\nu_e\nu_e}-
a_{\nu_X\nu_X})j_z=0\eqno(24)$$ is satisfied with the transition
width
$$\delta N_e(\nu_e\nu_X)\sim[N_e(\nu_e\nu_X)-4\pi(\sqrt{2}G_F)^{-1}
(a_{\nu_e\nu_e}-a_{\nu_X\nu_X})j_z]\tan2\theta_{\nu},\eqno(25)$$
where $N_e(\nu_e\nu_X)$ is an electron density at which the
resonance takes place; \\
(ii) $\nu_{eL}\to{\overline\nu}_{XL}$ is the resonance with flavor
and spin flipping which occurs at the condition
$$\Sigma_{\nu_{eL}\to{\overline\nu}_{XL}}=
2\delta^{12}_c+V_{eL}+V_{XL}+4\pi
(a_{\nu_e\nu_e}+a_{\nu_X\nu_X})j_z-\dot{\Phi}=0\eqno(26)$$ with
the resonance transition width
$$\delta N_e(\nu_e{\overline\nu}_X)\sim\frac{2\mu_{\nu_e{\overline\nu}
_X}B_{\bot}N_e(\nu_e{\overline\nu}_X)}{2\delta^{12}_c+4\pi
(a_{\nu_e\nu_e}+ a_{\nu_X\nu_X})j_z-\dot{\Phi}}.\eqno(27)$$

When the neutrino is the Dirac particles the resonance conversion
$\nu_{eL}\longrightarrow{\overline\nu}_{eL}$ takes place in
addition to above mentioned ones. It occurs at the condition
$$\Sigma_{\nu_{eL}\to{\overline\nu}_{eL}}=
2V_{eL}+4\pi(a_{\nu_e\nu_e}-a_{\overline{\nu}_e
\overline{\nu}_e}j_z-\dot{\Phi}=0.\eqno(28)$$

Further on, for the sake of simplicity, we assume that the
resonance localization places are situated rather far from one
another, that is, the following conditions hold
$$ N_e(k)+\delta N_e(k)< N_e(i)-\delta N_e(i),\eqno(29)$$
where $i,k=\nu_e\nu_X,\nu_e{\overline{\nu}}_X,
\nu_e{\overline{\nu}}_e.$ Now we may consider them as independent
ones. Then transition probabilities on resonances are given by the
expression
$${\cal D}_i=\exp{\{-{\gamma}^i (z_i)F_i\}},\eqno(30)$$
where ${\gamma}^i (z)$  is the adiabaticity parameter of
i-resonance, $z_i$ is the z-coordinate of i-resonance, and the
$F_i$ value depends on a kind of a resonance. In the most general
case, $F_i$ is dictated by the behavior of such quantities as
$\dot{\Phi}(z), V_{lL}$ and $j_z$ near the resonance. Assuming,
that all these quantities are linear functions on $z$, we get
$F_i=\pi/4$. It could be shown that the adiabaticity parameters
are determined by the relations
$$\gamma^i(z)={8({\cal H}_i)^2\over\sin^32\theta_i\Bigg|
{\displaystyle{d\over dz}\Sigma_i\Bigg|}},\eqno(31)$$ where
$$\sin^2{2\theta_i}=\frac{2{\cal
H}^2_i}{\Sigma^2_i+2{\cal H}^2_i},$$ and ${\cal H}_i$  is a
non-diagonal element of Hamiltonian in Eq.(5), corresponding to an
i-resonance transition.

\section{Conclusion}

The evolution equation for the electron neutrino flux moving in
the Sun is investigated. Our consideration carries general
character, that is, it holds for any SM extensions with massive
neutrinos. We assume that the neutrino possesses both dipole
magnetic and anapole moments while the solar magnetic field has
twisting nature. The resonance transitions of the electron
neutrino flux are found. For Dirac neutrinos these transitions are
as follows: (i) $\nu_{eL}\to\nu_{XL}$ (MSW resonance); (ii)
$\nu_{eL}\to{\overline\nu}_{XL}$ ($X=\mu,\tau$); (iii)
$\nu_{eL}\to{\overline\nu}_{eL}$. It should be stressed that in
the minimally extended SM (MESM) the two resonances last mentioned
have zero resonance transition widths and, as a result, they are
unobservable. When neutrinos are Majorana particles we may detect
only two resonance conversions: (i) $\nu_{eL}\to\nu_{XL}$; (ii)
$\nu_{eL}\to{\overline\nu}_{XL}$. Again, within the MESM the
latter has to be absent.

The MSW resonance may occur before the convective zone while
$\nu_{eL}\to{\overline\nu}_{XL}$- and
$\nu_{eL}\to{\overline\nu}_{eL}$-resonances could take place only
at upper layer of solar atmosphere in the sufficiently intensive
magnetic field. If the hypothesis of the $\nu_e$-induced decays
(H$\nu_e$ID) is the case, that is, decreasing the beta decay rates
is really caused by reduction of the solar neutrino flux, then it
is reasonable to suggest that $\nu_{eL}\to{\overline\nu}_{XL}$-
and $\nu_{eL}\to{\overline\nu}_{eL}$-resonances happen strictly
during the solar flare (SF).

After leaving from the solar surface the neutrino flux flies
150,000,000 km in a vacuum before it will reach the Earth. As this
takes place, weakening the electron neutrino flux is motivated by
the vacuum oscillations. It is well to bear in mind that vacuum
oscillations lead solely to $\nu_{eL}\to\nu_{XL}$ transitions. So,
when the SF is absent a terrestrial detector records the electron
neutrino flux weakened at the cost both of vacuum oscillations and
of the MSW resonance conversion. On the other hand, the electron
neutrino flux passed the SF region in preflare period proves to be
further weakened in so far as it undergoes one (Majorana neutrino)
or two (Dirac neutrino) additional resonance conversions, apart
from the MSW resonance and vacuum oscillations. It should be
particularly emphasized that the above mentioned statement
contradicts forecasts of the MESM and its confirmation will demand
revision of this SM extension.

Note, that correlations between nuclear decay rates and the
annually changing Earth-Sun distance reported for the first time
in Ref. \cite{JNJ09} could be also explained by the H$\nu_e$ID.
But there, the $\nu_{eL}$ flux reduction is caused by the vacuum
oscillations only.

Of course, establishing reasons of the $\nu_e$-induced decays is
one of the basic task of the contemporary physics which is so far
from the ultimate answer. However, closeness of the typical solar
neutrino energy and the nuclear binding energy per nucleon
suggests following simple mechanism. Since a neutrino does not
participate in strong interaction and has not electrical charge
the bulk of the solar neutrinos penetrate unobstructed to nucleus.
In so doing, neutrinos are not absorbed, while having given up a
part of energy they pass through the nucleus. As a result, the
decays of some elements of the periodic table become to be energy
allowed. Therefore, if the H$\nu_e$ID is true, then we may state:
some elements we belief that they are natural radioactive, in
actuality, are artificial radioactive because of the solar
neutrino flux bombardment.

Another consequence of the H$\nu_e$ID implies that nuclides with
the $\nu_e$-induced radioactivity could serve as real-time
neutrino detectors. Of course, each of them possesses definite
sensitivity relative to the variation of the solar neutrino flux.
Therefore, we have to find the nuclide having the maximum
sensitivity and use it to expand our understanding of both
neutrino physics and solar dynamics.

It should be stressed that appearance of $\overline{\nu}_{XL}$ and
$\overline{\nu}_{eL}$ in the solar neutrino flux could be detected
with the neutrino telescopes as well.

In order to certainly prove the H$\nu_e$ID we must obtain the
positive results using the well-controlled collider neutrino flux
for irradiation of the radioactive samples.

\end{document}